\documentclass[12pt]{iopart}

\usepackage{citesort}
\usepackage{graphicx,amssymb}
\usepackage{rotating}

\def\be{\begin{equation}}
\def\ee{\end{equation}}
\def\epem{$e^+e^-$\ }

\newcommand\prd{{Phys. Rev. D}}
\newcommand\aap{{A\&A}}%

\newcommand\aj{{AJ}}%
\newcommand\aaps{{A\&AS}}%
\newcommand\apss{{Ap\&SS}}%
\newcommand\apjl{{ApJ}}%
\newcommand\apj{{ApJ}}%
\newcommand\nat{{Nature}}%

\begin{document}


\title[Gendelev, Profumo \& Dormody: \epem from Fermi Gamma-Ray Pulsars]{The Contribution of Fermi Gamma-Ray Pulsars to \mbox{the local Flux of Cosmic-Ray Electrons and Positrons}}

\author{Leo Gendelev}
\address{Department of Physics, University of California, Santa Cruz, CA 95064, USA\\
Email:  {\tt  lgendele@ucsc.edu}}
\author{Stefano Profumo and Michael Dormody}
\address{Santa Cruz Institute for Particle Physics and Department of Physics,\\ University of California, Santa Cruz, CA 95064, USA\\
Email:  {\tt profumo@scipp.ucsc.edu, dormody@scipp.ucsc.edu}}

\begin{abstract}
We analyze the contribution of gamma-ray pulsars from the first Fermi-Large Area Telescope (LAT) catalogue to the local flux of cosmic-ray electrons and positrons ($e^+e^-$). We present new distance estimates for all Fermi gamma-ray pulsars, based on the measured gamma-ray flux and pulse shape. We then estimate the contribution of gamma-ray pulsars to the local \epem flux, in the context of a simple model for the pulsar \epem emission. We find that 10 of the Fermi pulsars potentially contribute significantly to the measured \epem flux in the energy range between 100 GeV and 1 TeV. Of the 10 pulsars, 2 are old EGRET gamma-ray pulsars, 2 pulsars were discovered with radio ephemerides, and 6 were discovered with the Fermi pulsar blind-search campaign. We argue that known radio pulsars fall in regions of parameter space where the \epem contribution is predicted to be typically much smaller than from those regions where Fermi-LAT pulsars exist. However, comparing the Fermi gamma-ray flux sensitivity to the regions of pulsar parameter space where a significant \epem contribution is predicted, we find that a few known radio pulsars that have not yet been detected by Fermi can also significantly contribute to the local \epem flux if (i) they are closer than 2 kpc, and if (ii) they have a characteristic age on the order of one mega-year.
\end{abstract}

\maketitle

\section{Introduction}

In the last two years, data on high-energy cosmic-ray electrons and positrons have attracted a great deal of interest. The positron fraction measurement reported by the PAMELA Collaboration in Ref.~\cite{Adriani:2008zr}, combined with the high-statistics electron-positron (\epem) flux measurement by the Fermi Large Area Telescope (LAT) Collaboration \cite{fermiepem} are widely believed to conclusively point to a new class of primary positron sources in the Galaxy (see e.g. \cite{grasso}). Interestingly, one such class that has been widely advocated and discussed of late is the pair-annihilation or the decay of Galactic particle dark matter (see e.g. \cite{Cirelli:2008pk}). A more mundane possibility, but by no means more or less disfavored by currently available data, is that the new class of positron sources is associated to nearby mature rotation-powered pulsars, as earlier envisioned in relation to the recent cosmic-ray data e.g. in Ref.~\cite{Hooper:2008kg,Yuksel:2008rf,Profumo:2008ms,Malyshev:2009tw,Kistler:2009wm}. 

While from the point of view of advancing our understanding of the fundamental structure of particle physics the dark matter scenario appears more appealing, pulsars might constitute if not the entirety of the detected positron excess, at least a potentially very significant background to searches for new physics. In this respect, gaining an accurate understanding of the contribution of pulsars to the local flux of high-energy \epem addresses questions that go beyond the realm of classical high-energy astrophysics and that are of relevance to the quest for physics beyond the Standard Model as well.

The Fermi-LAT Collaboration recently delivered \cite{psrcat} the first catalogue of 46 high-confidence gamma-ray pulsars, based on the first six months of LAT data. Roughly half of these pulsars were discovered using radio ephemerides, i.e. monitoring the locations and timing, in the gamma-ray sky, of known radio pulsars. The remaining pulsars were discovered by looking for a pulsed emission in bright gamma-ray sources, either corresponding to the potential locations of neutron stars identified via observations at other wavelengths, or via the pulsar blind-search campaign (see Ref.~\cite{Ziegler:2007zzb,blind,pulsarsinprep}); finally, 6 pulsars correspond to previously known EGRET gamma-ray pulsars \cite{Thompson:1994sg,Fierro:1995ny}.

The scope of the present study is to assess the contribution of pulsars in the first Fermi-LAT gamma-ray pulsar catalogue to the local flux of high-energy cosmic-ray electrons and positrons. There are several reasons to believe that gamma-ray pulsars contribute significantly to the local \epem flux. On theoretical grounds, the mechanism through which gamma rays are produced in the pulsar magnetosphere necessarily requires the generation of high-energy electron-positron pairs, independent of whether particle acceleration and radiation occurs in the neutron star polar cap (see e.g. Ref.~\cite{1975ApJ...196...51R,1983ApJ...266..215A,1982ApJ...252..337D,1996A&AS..120C.107D,1985Ap&SS.109..365R,1995AuJPh..48..571U}) or in the so-called outer gap (\cite{1986ApJ...300..500C,1996ApJ...470..469R,2002BASI...30..193H}). A possible caveat is the case of gamma rays produced by curvature emission from primary out-flowing electrons escaping magnetized neutron stars: in this case no positron counterpart would be produced by the gamma-ray emitting object. 

While over 1,800 radio pulsars are known \cite{2005AJ....129.1993M}, it is widely believed that these radio-loud objects only constitute a fraction of the wider class of rotation powered pulsars. The pulsar radio catalogue completeness is not only flux-limited, but -- more importantly -- it is limited by the pulsar radio beam geometry: a radio pulsar is visible only if the relatively narrow radio beam crosses the observer's line of sight. Gamma-ray pulsars, on the other hand, are believed to exhibit much wider beams: powerful nearby \epem injectors might thus very well be radio-quiet due to geometry, and yet produce a significant flux of gamma radiation. An outstanding example is the Geminga pulsar, that has been advocated e.g. in Ref.~\cite{Yuksel:2008rf,Hooper:2008kg} as the possible main contributor to the excess of cosmic-ray positrons detected by PAMELA. We therefore find it compelling to use the recent high-confidence list of gamma-ray pulsars to study the contribution of this class of primary positron emitters to the local \epem flux.

One of the main difficulties in assessing the contribution of gamma-ray pulsars to the local \epem cosmic-ray flux is the lack of information on the pulsar distance, especially if the pulsar has not been detected yet at other wavelengths. In the present analysis, we start, in Sec.~\ref{sec:dist}, from a systematic evaluation of the distance to the relevant gamma-ray pulsars in the first Fermi-LAT pulsar catalogue, calibrated on previously available information on distances to selected objects, and making  use of information from the measured gamma-ray flux and pulse shape and from the simulation results of the gamma-ray pulsar ``Atlas'' of Ref.~\cite{watters}. We then proceed in Sec.~\ref{sec:contrib} to evaluate the contribution from the Fermi-LAT gamma-ray pulsars to the flux of local leptonic cosmic rays, in the context of a simplified model for the flux of \epem injected in the ISM from the pulsar magnetosphere, and we compare with existing recent related studies \cite{Hooper:2008kg,Yuksel:2008rf,Profumo:2008ms,Malyshev:2009tw,Kistler:2009wm}. We then explore the relevant pulsar parameter space, with respect to the expected contribution from radio-quiet versus radio-loud pulsars (Sec.~\ref{sec:paramspace}), summarize and conclude (Sec.~\ref{sec:concl}).

\section{Gamma-ray pulsars: Distance Estimates}\label{sec:dist}

A pulsar is believed to consist of a rapidly rotating neutron star radiating
across different wavelengths. Pulsars are remarkably stable clocks, emitting
radiation with very precise time periods, with little or no variablity. While first
discovered at radio wavelengths, pulsars have been detected across all
wavelengths, including in the high-energy regime. With the launch of the space-based gamma-ray telescopes
EGRET and of the current generation instrument, the LAT on board the Fermi Space Telescope, pulsars have been discovered to pulse in gamma-rays, a crucial step to pinpoint the mechanism itself by which pulsars radiate.

Gamma-ray pulsars are rotation-powered pulsars, where the electromagnetic
radiation is fueled through the loss of angular momentum. While only a tiny (of the order of $10^{-6}$) fraction of
rotational energy is emitted at radio frequencies, the fraction of spin-down luminosity converted into gamma-rays is typically much larger, and not infrequently of order unity, making pulsar rotation
directly responsible for their high-energy radiation (see e.g. Ref.~\cite{psrcat,watters}). Broadly speaking, gamma-ray pulsars exist in two distinct populations: young, energetic pulsars, and old, recycled millisecond
pulsars. Since \epem are injected in the ISM in the early stage of a pulsar's lifetime, and since the very efficient energy losses for high-energy \epem limits their injection age, millisecond pulsars are not relevant for our present purposes, and we thus only focus here on
energetic, relatively young (characteristic age less than 10 Myr) pulsars. These pulsars spin with typical frequencies in the range 0.25-100 Hz and feature relatively large
spin-down energies $\dot E = 10^{34} - 10^{36}~\rm{erg}~\rm{s}^{-1}$. Electrons and positrons produced in the pulsars' magnetosphere, however, are trapped by the termination shocks of the pulsar wind neabula and flow into the ISM only when the envelope is suppressed enough. This process of diffusion of the \epem produced in the regions around the neutron star into the ISM has a characteristic timescale ranging between $10^4$ and $10^5$ yr (see e.g. \cite{1996ApJ...459L..83C}). Cosmic-ray \epem from very young pulsars, like Vela, might therefore not have diffused into the ISM yet \cite{Kistler:2009wm} and should not contribute significantly to the detected \epem local flux. The timescale of \epem injection in the ISM therefore generically further cuts out the youngest pulsars as contributors to the local \epem flux.

The detection of gamma-ray pulsars proceeds via two possible pathways \cite{psrcat}: (i) by using
the localization and timing information provided by radio ephemerides and folding on the gamma-ray photon data to detect gamma-ray
pulsations (known as \emph{radio-selected gamma-ray pulsars}), and
(ii) by calculating periodicity in the gamma-ray photon data alone using
blind frequency searches (known as \emph{gamma-selected gamma-ray
  pulsars} \cite{blind}). Both methods are robust and have yielded the discovery of close to 50
high-significance gamma-ray pulsars in total \cite{psrcat}.

The spin-down luminosity $\dot E$ of a rotation powered pulsar is simply the derivative of
the rotational kinetic energy $E = \frac{1}{2} I \omega^2$, $\dot E =
I \omega \dot \omega$. The characteristic age $t_{\rm ch}$ is the time it would
take an ideal magnetic dipole born spinning at infinite frequency to
slow to the observed pulsar frequency today. It is
only a rough estimate of the actual age of the pulsar, but it provides  useful information
in determining which population of pulsars might contribute the most to the
\epem local flux, since as we shall see below the injection time crucially enters the diffusive propagation of charged leptonic cosmic rays. The characteristic age can be calculated simply from the spin
parameters $\omega$ and $\dot \omega$: $t_{\rm ch} = - {\omega}/({2 \dot{\omega}})$.

To calculate the distances of the gamma-ray pulsars detected by Fermi-LAT, we use the heuristic model of Ref.~\cite{watters},
\begin{equation}\label{eq:grlum}
L_\gamma\approx w \dot E\approx C\times \left(\frac{\dot E}{10^{33}\ {\rm erg}\ {\rm s}^{-1}}\right)^{1/2}\times 10^{33}\ {\rm erg}\ {\rm s}^{-1},
\end{equation}
where $L_\gamma$ is the gamma-ray luminosity in the 0.1 GeV to 100 GeV energy range, $\dot E$ is the spin-down luminosity of the pulsar, $w\propto \dot E^{-1/2}$ is the gamma-ray efficiency, and $C$ is a constant of order unity that is in principle theoretically calculable given a pulsar gamma-ray production mechanism. Here, we take a phenomenological approach and we calibrate the coefficient $C$ by fitting it to Fermi-LAT gamma-ray data for the 15 pulsars from the brightest LAT source list catalogue \cite{brightsrc} with solid associated distance determinations (the latter taken from Ref.~\cite{psrcat}). We also considered the theoretically naive possibility that the gamma-ray luminosity is simply proportional to the available rotational energy of the neutron star, i.e. $L_\gamma\propto \dot E$, although somewhat disfavored by data from the above mentioned gamma-ray pulsars with distance determinations \cite{psrcat}. In this case, the gamma-ray luminosity of pulsars with a low spin-down luminosity is suppressed, leading in turn to a suppressed estimate for the distance to most of the relevant gamma-ray pulsars. This implies therefore that the assumption that $L_\gamma\propto \dot E$ leads to an over-estimate of the \epem contribution from gamma-ray pulsars. In other words, the assumption we employ for the gamma-ray pulsar luminosity leads to conservative predictions for the flux of cosmic-ray \epem.

The most precise distances to pulsars are determined from parallax measurements, but this method is only reliable for small distances ($< 500$ pc). For pulsars with no or unreliable parallax data, one must make use of other means of distance determination, such as radio Disperson Measure or X-ray absorption. In particular, the Dispersion Measure (DM) distance estimate hinges upon the effect on the pulsar radio beam of the integrated column density of free electrons between the Earth and the pulsar. The interaction between radio pulses and the charged particles (i.e. free electrons) in the Galaxy causes the pulses to be delayed by an amount depending on the electron density and on the radio frequency. By measuring the frequency shift and using a model for the distribution of free electrons, one can assess the distance to radio pulsars. Most radio pulsar distances in the Fermi Pulsar Catalog \cite{psrcat} are obtained in this way. Dispersion measurements in Ref.~\cite{psrcat} make use of the NE2001  electron density distribution model of Ref.\cite{Cordes:2003ik}. Other distance measurements used in the first LAT pulsar catalogue include using the Doppler shift of absorption or emission lines in the neutral hydrogen spectrum (HI), and X-ray flux measurements (see Ref.~\cite{psrcat}).

To estimate the distance to a pulsar with gamma-ray measurements, we use the relation \cite{watters}
\begin{equation}
L_\gamma=f_\Omega(4\pi  D^2) G_\gamma, 
\end{equation}
where $f_\Omega$ is the {\em flux correction factor}, which is a pulsar-dependent and model-dependent quantity that translates between the phase-averaged flux observed along the Earth line of sight and the gamma-ray flux averaged over the entire sky (see Ref.~\cite{Harding:2007ug}), $D$ is the estimated distance to the pulsar, and $G_\gamma$ is the $\gamma$-ray energy flux inferred by the Fermi analysis of the gamma-ray flux and spectra \cite{psrcat} (this quantity is indicated in \cite{psrcat} with the symbol $G_{100}$). 

The coefficient $C$ is assumed here to be universal, and to estimate its value we restrict our analysis to pulsars closer than 10 kpc and with a characteristic age $t_{\rm ch}<10^7$ yr. Incidentally, this implies that the spin-down luminosity of the pulsars we consider have spin-down luminosities $\dot E>0.5\times 10^{34}$ erg s$^{-1}$. While for very distant pulsars the uncertainty on the distance from radio measurements is very significant, the constraint on age stems from the fact that the maximal energy \epem produced $10^7$ yr ago can have today is smaller than roughly 20 GeV, and therefore pulsars older than $10^7$ year cannot contribute significantly to the local \epem flux in the energy range of interest here. Notice also that this cut on the pulsar age excludes milli-second pulsars as well. Interestingly, the range of spin-down luminosities selected by the cuts we employ corresponds to one where the heuristic formula of Eq.~(\ref{eq:grlum}) should hold: in several motivated pulsar gamma-ray emission models, including the slot and outer-gap frameworks, below a spin-down luminosity of $10^{34}$ erg s$^{-1}$ the gap saturates and the potential drop is almost entirely employed to fuel the pair cascade processes, breaking the relation in Eq.~(\ref{eq:grlum}), as pointed out e.g. in Ref.~\cite{Muslimov:2003yz,Zhang:2004bu}.  The constraints on age and distance leave 10 pulsars out of the 15 LAT bright pulsars with solid distance determinations, namely J0534+2200 (Crab), J0835-4510 (Vela), J0633+1746 (Geminga), J1709-4429, J1952+3252, J1057-5226, J1048-5832, J0631+1036, J1028-5819 and J1509-5850. 

The value for the flux correction factor $f_\Omega$ is calculated for each individual pulsar, and derived using the results of the ``Atlas'' presented in \cite{watters}, and making use of predictions from both the Outer Gap (OG) and the Two-Pole Caustic (TPC) models. The correction factor also depends on the magnetic inclination angle $\zeta$ and on the observer viewing angles $\alpha$; Since we do not have exact values for either one of these angles, we can at best find a range of values for $f_\Omega$ from the peak separation observed and reported in the first LAT pulsar catalogue, Ref.~ \cite{psrcat}. The first step is to choose the most plausible model for $f_\Omega$: Following the results of Ref.~\cite{watters}, for those pulsars which have a peak separation of 0.15-0.30 the OG model is the clear choice, while for separations larger than 0.47 the TPC model is typically more adequate. For intermediate peak separations, we find
that the OG model predicts values of $f_\Omega$ giving distance estimates closer to the catalogue distances in almost all cases. To estimate the coefficient $C$ we therefore employ, for intermediate peak separations, the $f_\Omega$ predictions from the OG model. 

Given the assumptions above for the pulsar model, we then estimate $f_\Omega$ for each of the 10 pulsars we use to calibrate the parameter $C$ in the following way: we derive the value of the gamma-ray efficiency $w$ from Eq.~(\ref{eq:grlum}) and use the results of the simulations of \cite{watters} corresponding to the value of $w$ closest to what we derive from the actual measured gamma-ray and spin-down luminosities. We then utilize the measured gamma-ray peak separation between the two most prominent peaks, as found in \cite{psrcat}, to record the range of viewing and magnetic inclination angles. Finally, this range is used to infer, again using the results of the simulated pulsars of Ref.~\cite{watters}, the flux correction factor $f_\Omega$ for each individual pulsar. Our final estimate for the parameter $C$ is $C=6.04\pm0.75$, where the quoted error results from the uncertainty on $f_\Omega$ and on $G_\gamma$ for the 10 pulsars we use to estimate the parameter.
  
With the determination of the best fit value for $C$, with the method to extract $f_\Omega$ outlined above, and with data on the measured gamma-ray energy fluxes and pulse shapes, we can obtain distance estimates for the Fermi-LAT gamma-ray pulsars of Ref.~\cite{psrcat}. In turn, these distances crucially enter the estimates of the local \epem flux produced by known gamma-ray pulsars. The uncertainties associated to the distance estimates stem from (i) the uncertainty on the Fermi-LAT measured gamma-ray energy flux $G_\gamma$, from (ii) the estimate of $f_\Omega$, the flux correction factor, and from (iii) the parameter $C$. We use standard error-propagation techniques to calculate the propagated uncertainty on the distance. The other quantity entering the gamma-ray pulsar distance estimates, namely the pulsar spin-down luminosity, is inferred from pulsar timing information and has virtually no associated uncertainty.


\begin{table*}[p]
\centering
\caption{A comparison of the distances quoted in the first Fermi-LAT pulsar catalogue \cite{psrcat} and our distance predictions based on the catalogue's gamma-ray data.\label{table1}}
\vspace{0.5cm}

{\small
\begin{tabular}{lcc}
\hline\hline
{Pulsar Name} & {Catalog Distance (kpc)} & {gamma-ray distance (kpc)} \\
\hline

J0357+32  &  \ldots  &  0.82  $\pm$  0.16\\
J1732--31  &  \ldots  &  1.52  $\pm$  0.31\\
J1741--2054  &  .38 $\pm$ .11  &  1.11  $\pm$  0.32\\
J1809--2332  &  1.7 $\pm$ 1  &  1.45  $\pm$  0.29\\
J1836+5925  &  $<$.8  &  0.44  $\pm$  0.4\\
J2021+4026  &  1.5 $\pm$ 0.45  &  0.44  $\pm$  0.06\\
J0633+1746   &  .25 $\pm$ .12  &  0.28  $\pm$  0.09\\
J0659+1414   &  .288 $\pm$ .033  &  1.39  $\pm$  0.17\\
J1057--5226  &  0.72  &  1.02  $\pm$  0.2\\
J2043+2740  &  1.8 $\pm$ .54  &  3.35  $\pm$  0.77\\

\hline

J1509−-5850  &  2.6 $\pm$ .8  &  2.87  $\pm$  0.56\\
J0631+1036  & 0.75-3.62   &  4.87  $\pm$  1.25\\
J1952+3252  &  2 $\pm$ .5  &  3.12  $\pm$  0.44\\
J0835--4510   &  .287 $\pm$ .019  &  0.29  $\pm$  0.02\\
J2032+4127  &  1.6-3.6  &  2.75  $\pm$  0.69\\
J1958+2846  &  \ldots  &  2.99  $\pm$  0.68\\
J2238+59  &  \ldots  &  3.99  $\pm$  0.94\\
J1907+06  &  \ldots  &  3.34  $\pm$  1.03\\
J1826--1256  &  \ldots  &  2.27  $\pm$  0.48\\
J1813--1246  &  \ldots  &  3.19  $\pm$  0.69\\
J1459--60  &  \ldots  &  2.93  $\pm$  0.76\\
J1418--6058  &  2.0-5.0  &  3.03  $\pm$  0.73\\
J0633+0632  &  \ldots  &  2.02  $\pm$  0.49\\
J0007+7303  &  1.4 $\pm$ .3  &  2.41  $\pm$  0.74\\
J0534+2200   &  2 $\pm$ .5  &  2  $\pm$  0.5\\
J1709--4429  &  1.4-3.6  &  2.5  $\pm$  1.11\\
J1048--5832  &  2.71 $\pm$ .81  &  2.71  $\pm$  0.82\\
J1028−-5819  &  2.33 $\pm$ .7  &  2.33  $\pm$  0.73\\
J2021+3651  &  2.1 $\pm$ 2.1  &  2.1  $\pm$  2.11\\
J2229+6114  &  0.8-6.5  &  3.65  $\pm$  2.85\\
\hline
\end{tabular}
}
\end{table*}

\begin{figure}[t]
\begin{center}
\includegraphics[width=12cm,clip]{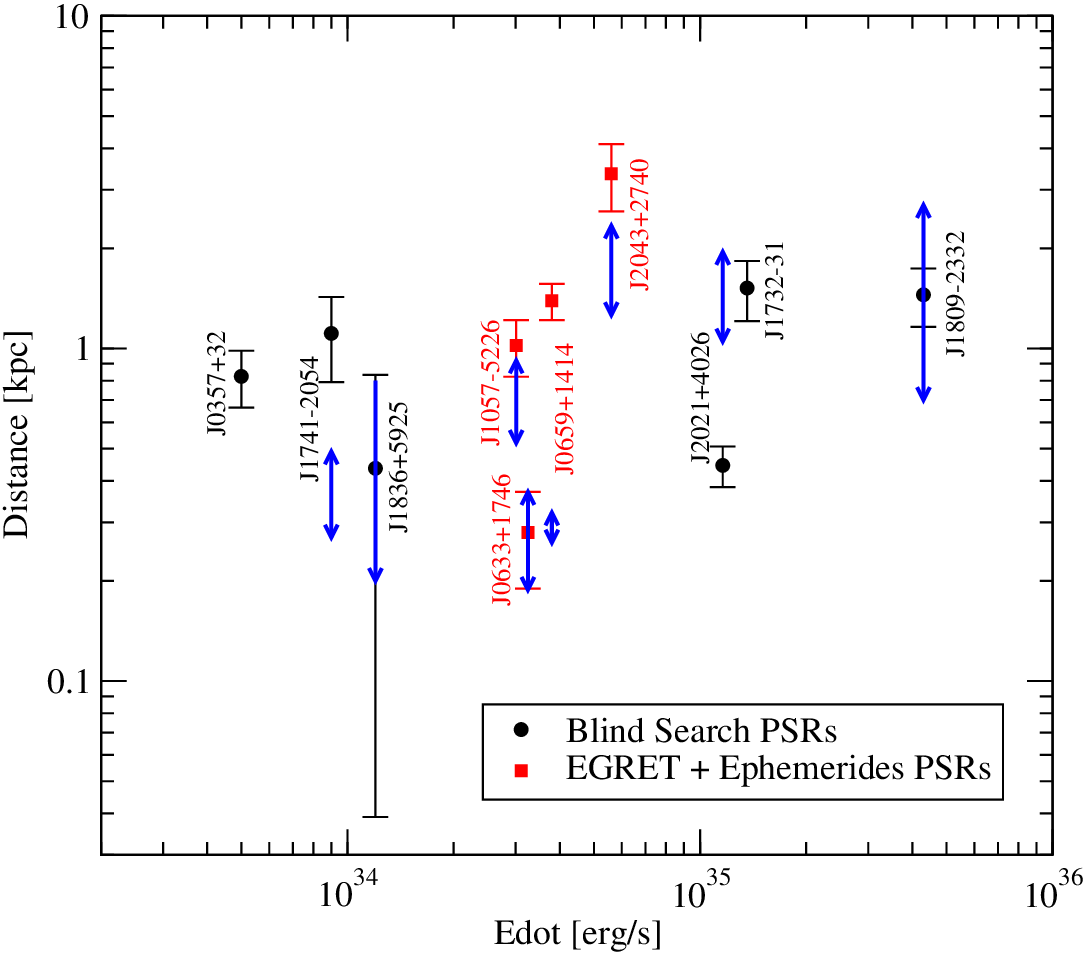}\\
\caption{A comparison of the pulsar distances calculated with Fermi-LAT gamma-ray flux data and of distances obtained, for some of the same pulsars, with other methods (indicated with a blue double arrow; see Ref.~\cite{psrcat}). Distances, in kpc, are plotted against the pulsars' spin-down luminosity $\dot E$ in erg $s^{-1}$. Black circles indicate pulsars discovered in the Fermi-LAT pulsar blind search \cite{blind}, while red squares refer to either EGRET pulsars \cite{egretpsr} or to pulsars found with radio ephemerides.  \label{fig:edotdistgr}}
\end{center}
\end{figure}

Tab.~\ref{table1} quotes the distances for the pulsars in the Fermi-LAT catalogue and compares them with our estimates based on the measured gamma-ray flux and gamma-ray pulse properties (the pulsar names are listed in the first column, the catalogue distances, when available, in the second column and our estimates in the third column). We include only pulsars with a characteristic age below 10 Myr (older pulsars do not contribute to the cosmic-ray \epem flux in the energy range we are interested in here) and nearby (i.e. distances less than 10 kpc) objects. The first ten pulsars in the table correspond to the top 10 pulsars contributing to the local flux of cosmic-ray \epem (we present details on how we estimate this flux in the next section). 

We show in Fig.~\ref{fig:edotdistgr} a visual comparison of the DM distances and gamma-ray distances, as a function of the pulsar's spin down luminosity $\dot E$, for those same 10 \epem-brightest pulsars. Objects indicated in black correspond to pulsars found with gamma-ray data, while those in red to either previously known gamma-ray pulsars (EGRET pulsars) or pulsars discovered with radio ephemerides. The thick blue double-arrowed lines indicate alternate distance estimates for the pulsars under considerations, for those pulsars with a distance determination quoted in Ref.~\cite{psrcat}.

We find that several of the distance estimates perfectly agree with the known DM pulsar distances from the catalog (e.g. J1836+5925, J1057-5226, J0633+1746, and J1809-2332; the distance estimates are compatible at the 2$\sigma$ level for J1741-2054 and for J2043+2740). Only in two cases do we find that the gamma-ray distance either overestimates (for J0659+1414) or underestimates (for J2021+4026) the DM distance as determined in the Fermi-LAT pulsar catalogue from the NE2001 model. In the case of PSR J2021+4026, the gamma-ray luminosity reported in the pulsar LAT catalogue \cite{psrcat} includes a bright un-pulsed emission, potentially associated to a secondary nearby source of gamma rays, or to magnetospheric emission \cite{psrcat}. The LAT catalogue \cite{psrcat} also points out PSR J0659+1414 as an outlier with an inferred luminosity a factor 30 below the general trend for other gamma-ray pulsars. One possibility for the apparent very low gamma-ray efficiency of PSR J0659+1414 is the smallness of the viewing angle $\zeta<20^\circ$, as envisioned e.g. in Ref.~\cite{Everett:2000yj}.

Under- and over-estimates of pulsar distances determined with gamma-ray data are therefore dominantly due to the pulsar-dependent gamma-ray conversion efficiency, which in our setup is assumed to essentially be universal and mirrored by the average value of the constant $C$ (although some pulsar-dependent information on the flux correction factor $f_\Omega$ is included in the present analysis). Nevertheless, the agreement we find between DM distances and gamma-ray distances for pulsars contributing significantly to the local \epem flux is remarkable and gives us confidence on the distance extrapolation for pulsars for which no DM estimate is known. Finally, we also point out that among pulsars that do not contribute significantly to the local \epem flux (lower part of Tab.~\ref{table1}), we also find very good agreement between our gamma-ray distance predictions and distance estimates as listed in Ref.~\cite{psrcat}. The distance estimates we obtain with gamma-ray data all agree with other distance determinations at the 2-$\sigma$ level or better.

\section{The Contribution of Gamma-ray pulsars to the Cosmic-Ray \epem Flux}\label{sec:contrib}

The \epem output from pulsars has been investigated recently in several studies (e.g. Ref.~\cite{Yuksel:2008rf,Hooper:2008kg,Profumo:2008ms,Kistler:2009wm}), given the interest in primary sources of energetic cosmic-ray positrons sparked by the PAMELA results on the positron fraction \cite{Adriani:2008zr}. The level of emission of high energy \epem is expected to be pulsar-dependent, and no consensus exists on a robust theoretical prediction for the spectrum and normalization of this flux. For the sake of illustration, we adopt here a simple and easily reproducible model for the \epem emission from pulsars, and for the propagation of high-energy \epem from the neutron star to the Earth. Also for simplicity and to allow for a comparison among the various pulsars we adopt the same parameters and assumptions for all objects. We closely follow here the notation and the analytical expressions of Ref.~\cite{grasso} and references therein.

We assume an \epem spectrum at source of the form
\begin{equation}
\label{eq:injspec}
Q(E,t,\vec r)=Q_0 \left(\frac{E}{1\ {\rm GeV}}\right)^{-\Gamma}\ \exp[-E/E_{\rm cut}]\delta(t-t_0)\delta(\vec r),
\end{equation}
and we take a spectral index $\Gamma=1.7$. This value is in the range of the gamma-ray spectral indexes reported in the Fermi-LAT pulsar catalogue \cite{psrcat}, where $1\lesssim\Gamma_\gamma\lesssim2$ with peaks at $\gamma\sim1.3$ and 1.7. Although the physical mechanism that produces \epem in the pulsar magnetosphere is presumably the same that then gives rise to the gamma-ray emission \cite{1987ICRC....2...92H}, reacceleration in the pulsar wind nebula (PWN) or in the supernova envelope can significantly affect the \epem spectral index. The value of $\Gamma\sim1.7$ we adopt here is however consistent with the \epem population plausibly responsible for the synchrotron and inverse Compton emission from e.g. the Crab PWN \cite{1988ApJ...327..853R} (see also the discussion in \cite{Kistler:2009wm}).

We adopt the value $E_{\rm cut}=1$ TeV, compatible with the expected decrease with time (see e.g. Ref.~\cite{Busching:2008zzb}) of the large cutoff energy expected from the observed very high-energy gamma-ray emission from the PWN of young pulsars \cite{1997MNRAS.291..162A}. We neglect here the time delay between the pulsar birth and the point in time when \epem are injected in the inter-stellar medium ($t_0$ in Eq.~(\ref{eq:injspec})) because (i) the uncertainty in the characteristic age versus the actual pulsar age is of the same order of magnitude as the ISM injection time scale, and (ii) our results are basically insensitive to this parameter for the pulsars that contribute the most to the local \epem flux.

Finally, the normalization parameter $Q_0$ was set for each pulsar to the value such that
\begin{equation}
\int_{m_e}^\infty\ E\times Q(E){\rm d}E=E_{\rm out}=\eta\frac{\dot E \ t^2_{\rm ch}}{\tau},\quad{\rm with} \ \tau\simeq10^4\ {\rm yr}\ {\rm and}\ \eta=0.4.
\end{equation}
with $t_{\rm ch}$ the characteristic pulsar age, $\tau$ the characteristic luminosity decay time, $\dot E$ the spin-down luminosity, and $\eta$ the \epem production efficiency (our results are easily linearly rescaled for different values of the last parameter). For the relatively large \epem production efficiency we assume here, $\eta=0.4$, the resulting total energy output in \epem $E_{e^+e^-}$ for the 10 pulsars we consider is always of the order of $10^{48}$ erg, with the exception of the older pulsars J2043+2740 (for which $E_{e^+e^-}\sim10^{50}$ erg) and of J1836+5925 and J1057-5226 (for which $E_{e^+e^-}\sim10^{49}$ erg). Observations of pulsar wind nebulae at radio, X-ray and very high-energy gamma-ray energies indicate however that $E_{e^+e^-}\gtrsim10^{48}$ erg \cite{Kistler:2009wm}, with uncertainties associated to the \epem diffusion and magnetic fields in the nebulae: such large total energy outputs appear therefore observationally plausible. 

To calculate the local \epem flux from a source term such as the one we adopt in  Eq.~(\ref{eq:injspec}) we consider the following standard cosmic-ray diffusion-loss transport equation:
\begin{equation}\label{eq:diffloss}
\frac{\partial N_e(E,t,\vec r)}{\partial t}-D(E)\nabla^2 N_e-\frac{\partial}{\partial E}(b(E)N_e)=Q(E,t,\vec r),
\end{equation}
where $D(E)=D_0(E/1\ {\rm GeV})^\delta$ is the rigidity-dependent diffusion coefficient, for which we assume the customary values $D_0=3.6\times 10^{28}\ {\rm cm}^2/s$ and $\delta=0.33$ \cite{grasso}, and where $b(E)=b_0 E^2$ is the energy loss term, which includes the dominant synchrotron and inverse Compton energy loss mechanisms, with the numerical value $b_0=1.4\times10^{-16}\ {\rm GeV}^{-1}{\rm s}^{-1}$. Analytic solutions exist for the above transport equation with the source term we take here (see e.g. \cite{1995PhRvD..52.3265A}), and the resulting contribution to the local \epem flux reads \cite{grasso}:
\begin{eqnarray}\label{eq:sol}
\nonumber N_e(E,t,\vec r)&=&\frac{Q_0}{\pi^{3/2} R^3_{\rm diff}(E,t)}\left(1-\frac{E}{E_{\rm max}(t)}\right)^{\Gamma-2}\left(\frac{E}{1\ {\rm GeV}}\right)^{-\Gamma}\\
&&\times\exp\Big[-\frac{E}{E_{\rm cut}}\frac{1}{1-E/E_{\rm max}}-\left(\frac{r}{R_{\rm diff}}\right)^2\Big]
\end{eqnarray}
with
\begin{equation}\label{eq:rdiff}
R_{\rm diff}(E,t)\simeq2\left(D(E)t\frac{1-(1-E/E_{\rm max})^{1-\delta}}{(1-\delta)E/E_{\rm max}}\right)^{1/2}
\end{equation}
and where the maximal energy, i.e. the energy an electron or positron injected with arbitrarily large energy would have after a time $t$, is $E_{\rm max}(t)=(b_0 t)^{-1}$.


\begin{sidewaystable*}[p]
\centering
\caption{Physical properties of the Fermi-LAT pulsars contributing to the local \epem flux, and the resulting relative contributions at 100 GeV (R100) and at the peak energy Epeak (Rpeak) with respect to the \epem flux measured by Fermi-LAT \cite{fermiepem}\label{table2}} 

\vspace{0.5cm}

{\footnotesize
\begin{tabular}{lccccccc}
\hline\hline
{Pulsar Name}  &  {Pulsar Type}  &  {Spectral Index} &  Age (kyr)  &  Edot ($10^{34}$ erg)  &  R100  &  RPeak    &  Epeak (GeV)  \\[1ex]
\hline

J0357+32  &  BS  &  1.29(0.22)  &   585.00  &  0.50  &  8.74E-03$_{-3.24E-03}^{+3.28E-03}$  &  1.43E-02$_{-3.71E-03}^{+4.06E-03}$  &  217$_{-7.4}^{+9.23}$ \\[1ex]
J1732-31  &  BS  &  1.27(0.14)  &    120.00  &  13.60  &  2.86E-05$_{-2.86E-05}^{+8.68E-03}$  &  5.15E-03$_{-7.65E-02}^{+5.04E-03}$  &  829.84$_{-187.37}^{+162.71}$ \\[1ex]
J1741-2054  &  BS  &  1.39(0.17)  &    392.10  &  0.90  &  5.32E-03$_{-4.62E-03}^{+1.14E-02}$  &  1.41E-02$_{-1.59E-02}^{+1.02E-02}$  &  303.34$_{-31.77}^{+37.48}$ \\[1ex]
J1809-2332  &  BS  &  1.52(0.07)  &    67.60  &  43.00  &  2.94E-07$_{-2.94E-07}^{+2.60E-03}$  &  1.79E-03$_{-8.65E-02}^{+1.79E-03}$  &  1262.48$_{-344.39}^{+301.68}$ \\[1ex]
J1836+5925  &  BS  &  1.35(0.04)  &    1800.00  &  1.20  &  4.79E-02$_{-1.15E-02}^{+1.93E-03}$  &  4.96E-02$_{-2.07E-03}^{+1.25E-02}$  &  91.04$_{-0.42}^{+2.56}$ \\[1ex]
J2021+4026  &  BS  &  1.79(0.04)  &    76.80  &  11.60  &  3.82E-02$_{-3.82E-02}^{+2.89E-02}$  &  1.61E-01$_{-1.86E-02}^{+1.61E-01}$  &  643.9$_{-44.65}^{+-633.9}$ \\[1ex]
J0633+1746   &  EGRET   &  1.08(0.02)    &  342.00  &  3.25 & 6.56E-02$_{-1.06E-02}^{+6.15E-03}$  &  1.17E-01$_{-7.01E-03}^{+1.30E-02}$  &  288.46$_{-3.56}^{+6.64}$ \\[1ex]
J0659+1414   &  Eph  &  2.37(0.5)  &    110.00  &  3.80  &  1.64E-05$_{-1.62E-05}^{+4.91E-04}$  &  2.07E-03$_{-8.46E-03}^{+1.80E-03}$  &  846.37$_{-114.11}^{+109.94}$ \\[1ex]
J1057-5226  &  EGRET  &  1.06(0.1)  &   535.00  &  3.01  &  3.56E-02$_{-1.93E-02}^{+2.54E-02}$  &  6.73E-02$_{-3.11E-02}^{+2.80E-02}$  &  237.77$_{-12.33}^{+14.21}$ \\[1ex]
J2043+2740  &  Eph  &  1.07(0.66)  &   1200.00  &  5.60  &  4.83E-03$_{-4.77E-03}^{+6.83E-02}$  &  9.35E-03$_{-8.24E-02}^{+9.00E-03}$  &  159.13$_{-23.17}^{+0}$ \\[1ex]

\hline
\end{tabular}
}
\end{sidewaystable*}

With the parameter assumptions outlined above, and with the distance estimates obtained in the preceding section, we obtain the results reported in Tab.~\ref{table2} for the 10 pulsars with the largest contributions to the local \epem flux (all other pulsars contribute less than 0.1\% to the measured \epem flux at the energies corresponding to their peak emission). In the table we indicate, together with the pulsar name, a series of physical properties, including whether the object in question was discovered via the pulsar blind search campaign, or via radio ephemerides or whether it was an already-known gamma-ray pulsar (i.e. detected by EGRET), the gamma-ray spectral index, the pulsar age and its spin-down luminosity. The last three columns indicate the relative contribution to the local \epem flux (as measured by Fermi-LAT \cite{fermiepem}) at an energy of $E_{e^\pm}=100$ GeV (R100) and at the energy Epeak where the flux times energy cubed peaks (Rpeak).

\begin{figure}[t]
\begin{center}
\includegraphics[width=14cm,clip]{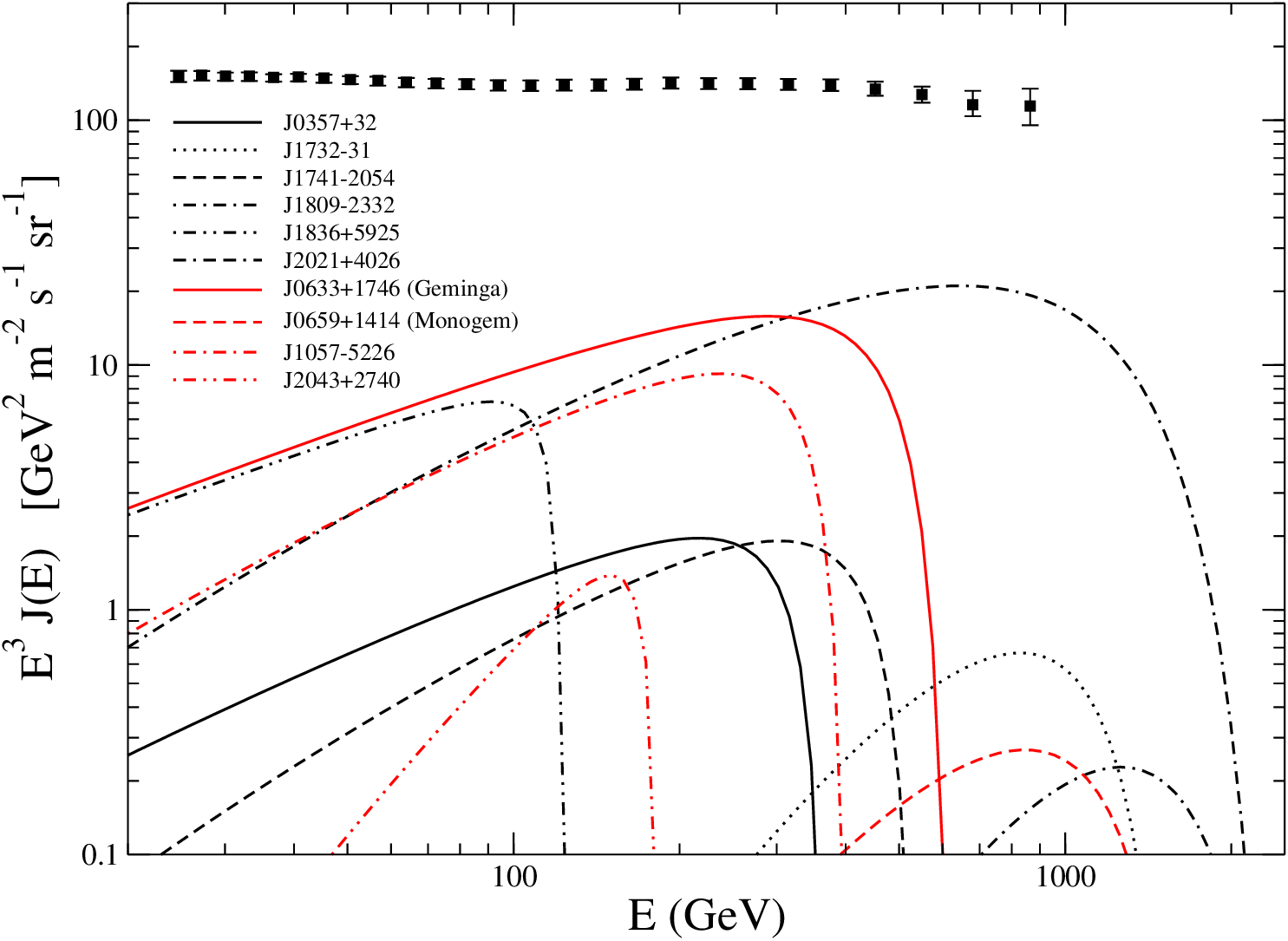}\\
\caption{The spectrum of the 10 Fermi-LAT pulsars giving the largest contributions to the local \epem flux, assuming an \epem injection effciency $\eta=0.4$ and an \epem spectral index $\Gamma=1.7$ with a cutoff $E_{\rm cut}=1$ TeV for all pulsars. Black lines refer to blind search gamma-ray selected pulsars, red lines to all other pulsars.  The data points reproduce the \epem spectrum measured by Fermi \cite{fermiepem} \label{fig:spectrum}}
\end{center}
\end{figure}

We illustrate the actual \epem flux predictions for the ten contributing pulsars in Fig.~\ref{fig:spectrum}. Red lines correspond to pre-Fermi ``known'' gamma-ray pulsars, including the Monogem (J0659+1414) and Geminga (J0633+1746) pulsars, while black lines refer to the new pulsars discovered with the blind search campaign on the Fermi-LAT data. We also show, for comparison, the Fermi-LAT measurement of the local \epem flux \cite{fermiepem}. From the figure, it is clear that bright gamma-ray pulsars can play an important role as contributors to the local flux of energetic \epem. A potentially very significant contribution stems from the gamma-ray selected pulsar J2021+4026.

While most of the pulsars we consider here in Tab.~\ref{table2} and Fig.~\ref{fig:spectrum} have never been previously studied as sources of cosmic-ray electrons and positrons, two significant and noteworthy exceptions (Geminga and Monogem) allow us to directly compare in detail with the results of recent related analyses \cite{Hooper:2008kg,Yuksel:2008rf,Profumo:2008ms,Malyshev:2009tw,Kistler:2009wm} and \cite{grasso}. In particular, our present setup follows exactly the same conventions as Ref.~\cite{Profumo:2008ms} and \cite{grasso}, so our results match with those presented there, modulo setting the same parameter values (most of the choices we make here in fact match those in \cite{grasso}, while a broader range of parameter choices is considered in \cite{Profumo:2008ms}). Ref.~\cite{Hooper:2008kg} considers both Monogem and Geminga, although with a different diffusion setup (especially for the rigidity dependence $\delta=0.6$ versus $\delta=0.33$ adopted here) and with a different estimate for the distance to Geminga. The range of injection spectral indices (1.5 -- 2) encompasses our choice of 1.7. Our estimate for the total energy output for Geminga is larger than what considered in \cite{Hooper:2008kg}, although the smaller distance partly compensate for it. The output we estimate for Monogem is instead in the middle of the range envisioned in \cite{Hooper:2008kg}. Factoring in the above mentioned differences, our results are in line with those presented in ref.~\cite{Hooper:2008kg} for the two considered pulsars. The spectra presented in ref.~\cite{Yuksel:2008rf} and \cite{Kistler:2009wm} for the Geminga pulsar appear to be slightly different from what we show in fig.~\ref{fig:spectrum}. This is due to both slightly different choices for the cosmic-ray propagation parameters (smaller diffusion coefficient, $\delta=0.4$), and to the fact that in \cite{Yuksel:2008rf} a time-dependence for the process of \epem injection in the ISM was included (leading to an output tail of higher energy cosmic rays), and that in \cite{Kistler:2009wm} a fully relativistic solution to the diffusion-loss equation based on the J\"uttner formalism is employed. In both cases, though, the overall predicted \epem output from Geminga is comparable to the present estimate based on gamma-ray data. Finally, Ref.~\cite{Malyshev:2009tw} uses a theoretical setup which is very close to what we employ here, but does not single out the Geminga or Monogem pulsars, hence a direct comparison with our results is not directly obvious. However, their results on the overall \epem output are once again in line with what we present here.

\begin{figure}[t]
\begin{center}
\includegraphics[width=16cm,clip]{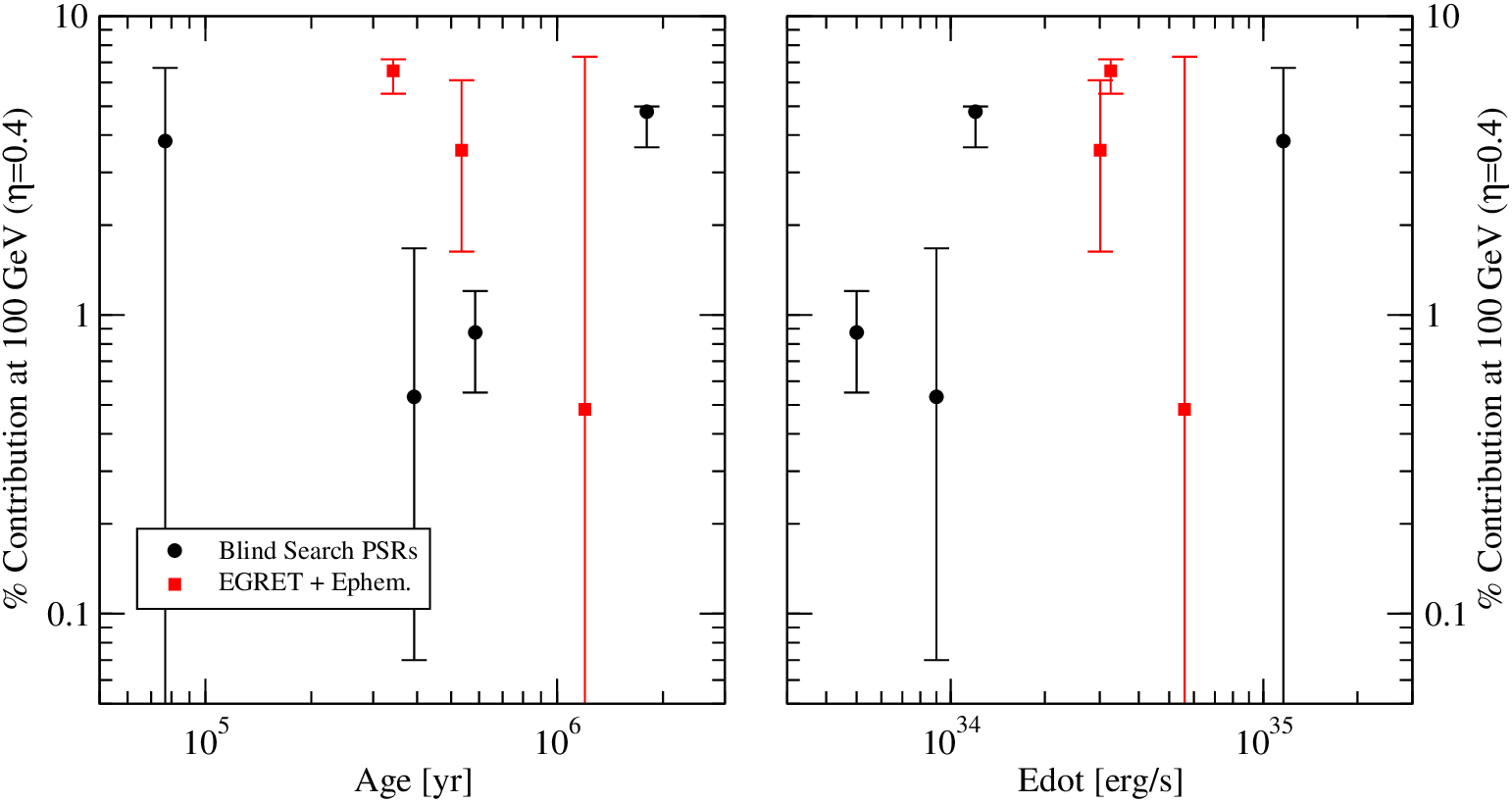}\\
\caption{Percent contribution to the total \epem flux, as measured by Fermi-LAT \cite{fermiepem}, at $E_{e^+e^-}=$100 GeV. In the left panel we show said contribution as a function of the pulsar age, on the right as a function of the pulsar spin-down luminosity. \label{fig:R100}}
\end{center}
\end{figure}

\begin{figure}[t]
\begin{center}
\includegraphics[width=16cm,clip]{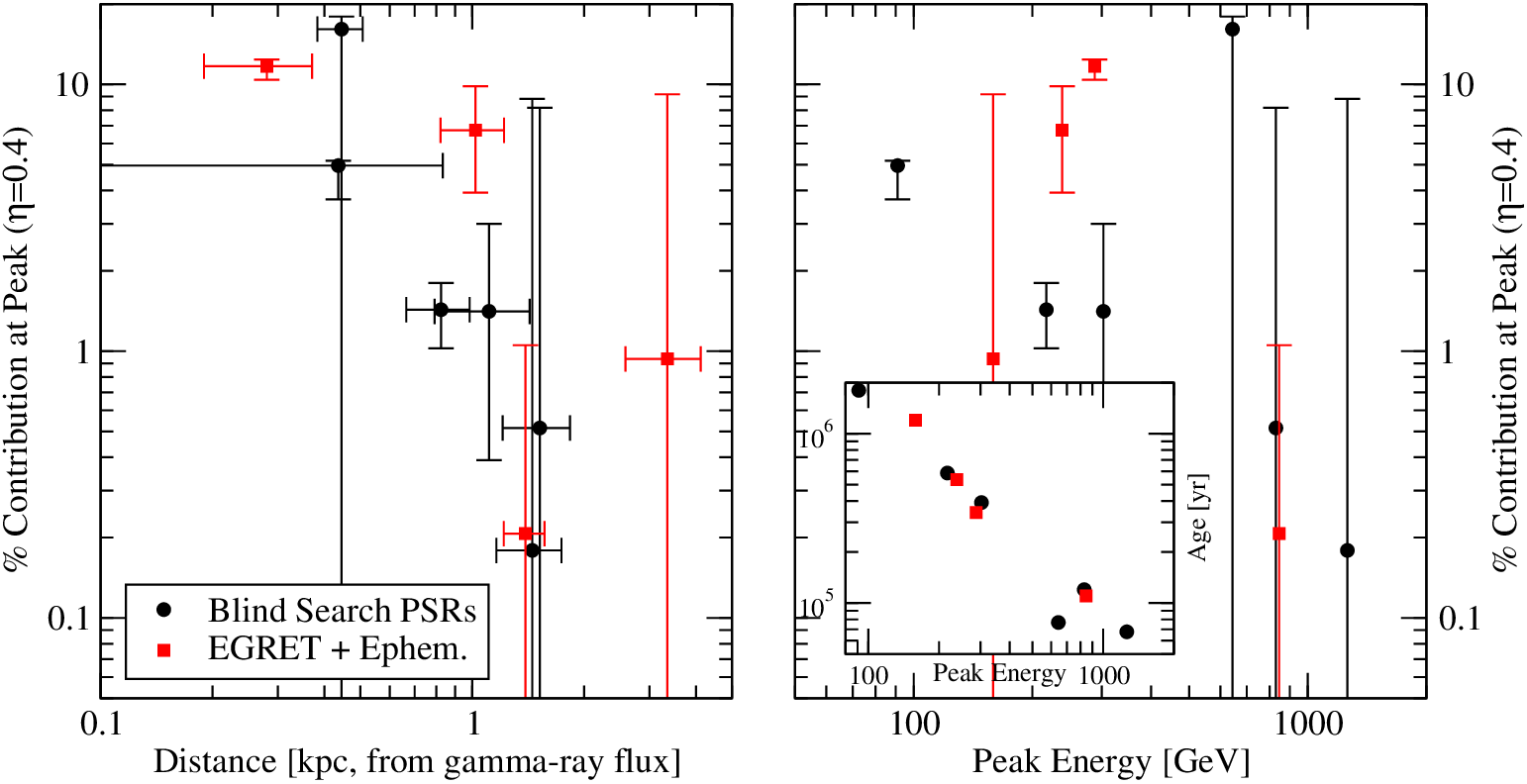}\\
\caption{Percent contribution to the total \epem flux, as measured by Fermi-LAT \cite{fermiepem}, at the peak energy for each individual pulsar. In the left panel we show said contribution as a function of the pulsar distance estimated from the Fermi-LAT measured gamma-ray flux (we also show errors on the distance estimate), on the right as a function of the \epem flux peak energy (see also Fig.~\ref{fig:spectrum}). The inset in the right panel shows the flux peak energy as a function of the pulsar age.\label{fig:Rpeak}}
\end{center}
\end{figure}

We further elaborate on this point in Fig.~\ref{fig:R100} and \ref{fig:Rpeak}. With the same color coding (black for gamma-ray selected pulsars, red for radio-selected and EGRET pulsars), we show in Fig.~\ref{fig:R100} the percent contribution of gamma-ray pulsars to the measured \epem flux at $E_{e^\pm}=100$ GeV, with the parameter space assumptions specified above, as a function of the pulsar age (left panel) and spin-down luminosity (right panel). The error bars mirror the uncertainty in the pulsar distance, as determined from our analysis based on gamma-ray luminosity and pulse shape. We find that at least four new Fermi-LAT pulsars should give a non-negligible contribution (more than 1\%) at $E_{e^\pm}=100$ GeV. We thus find that the currently known gamma-ray pulsars that might contribute to the local \epem flux span a wide range of both characteristic age and spin-down luminosities, respectively between $10^5$ and $2\times 10^6$ yr and $\dot E$ between $0.5\times 10^{34}$ and $10^{35}$ erg/s.

The relative contribution from pulsars to the \epem at the peak of $E_{e^\pm}^3 J$ ($J$ being the local \epem flux per steradian from the given pulsar) is shown in Fig.~\ref{fig:Rpeak}, as a function of the pulsar distance and peak location. In the figure we indicate the horizontal errorbars corresponding to the uncertainty on our gamma-ray pulsar distance determination, as well as the resulting propagated uncertainty on the contribution at peak. As expected, the main pulsar \epem contributors lie relatively close to us, at most at a distance of 1 kpc. The contribution from further pulsars decreases steeply with distance.

The energy where the pulsar contribution to the \epem flux is expected to peak, with respect to the measured local \epem cosmic-ray flux, ranges between 100 GeV and 1 TeV. The location of this peak is a function of the pulsar age, that sets the maximal energy $E_{\rm max}(t)$ electrons and positrons can have a time $t$ after injection. This is illustrated in the inset of Fig.~\ref{fig:Rpeak}, right panel, where we correlate the peak position to the pulsar age. The contribution from older pulsars peaks at lower frequencies, while younger pulsars tend to contribute at higher energies.


\section{Can Gamma-Ray Quiet Pulsars Significantly Contribute to the Local \epem Flux?}\label{sec:paramspace}

While the ratio of the gamma-ray flux from a given pulsar over the pulsar luminosity scales as $\sim 1/D^2$, i.e. it is inversely proportional to the square of the distance of the observer to the pulsar, the contribution of pulsars to the local \epem flux has a significantly more complicated dependence on distance. This stems from the fact that the propagation of \epem from the pulsar to the Earth proceeds via a diffusive process. Similarly, unlike the gamma-ray flux, the \epem flux has a critical dependence on the pulsar age. Here, as detailed above, we model the diffusive mechanism of charged particles losing energy and changing direction scattering off of the magnetic fields irregularities in the picture encompassed by Eq.~(\ref{eq:diffloss}). In turn, this implies the non-trivial and highly non-linear dependence of the pulsar contribution to the \epem flux on distance given in Eq.~(\ref{eq:sol}). In particular, this dependence is very sensitive to the pulsar age and to the assumed diffusion coefficient, via the energy-dependent diffusion radius $R_{\rm diff}$.

As a result of the non-trivial propagation of \epem in the Galaxy, it is not obvious which of the bright gamma-ray pulsars are the most important contributors to the local cosmic-ray \epem flux. In this section we investigate and outline the portions of the pulsar parameter space where pulsars are expected to contribute to the \epem flux, and the overlap of these regions with the Fermi-LAT sensitivity to the emission from gamma-ray pulsars. This allows us to determine which regions of the pulsar parameter space, important to understand the local \epem flux, might be within reach with further improvements to the Fermi-LAT sensitivity to pulsed gamma-ray emission, and which regions are equally important but beyond the reach of the LAT.

\begin{figure}[t]
\begin{center}
\mbox{\includegraphics[width=8cm,clip]{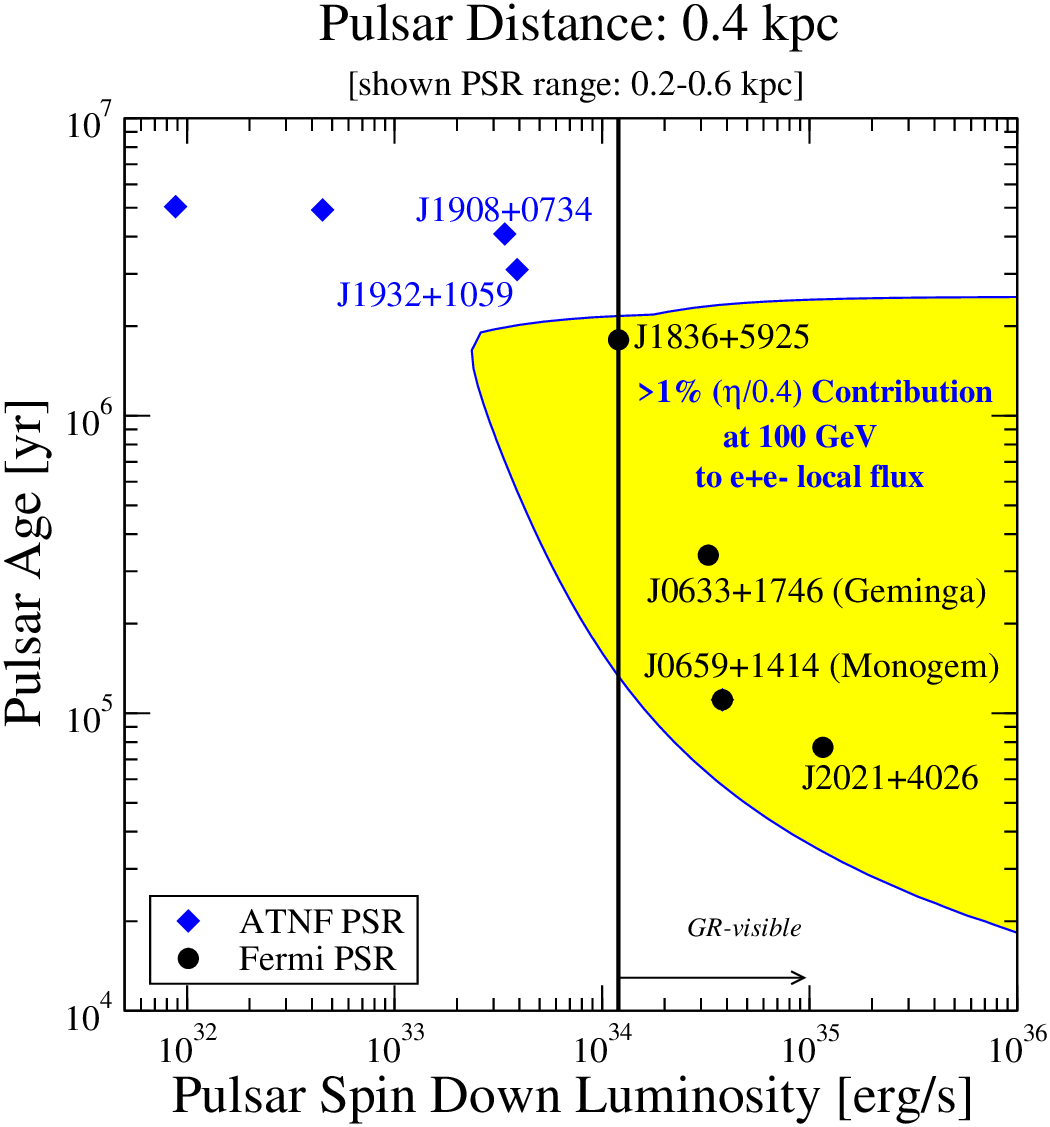}\qquad \includegraphics[width=8cm,clip]{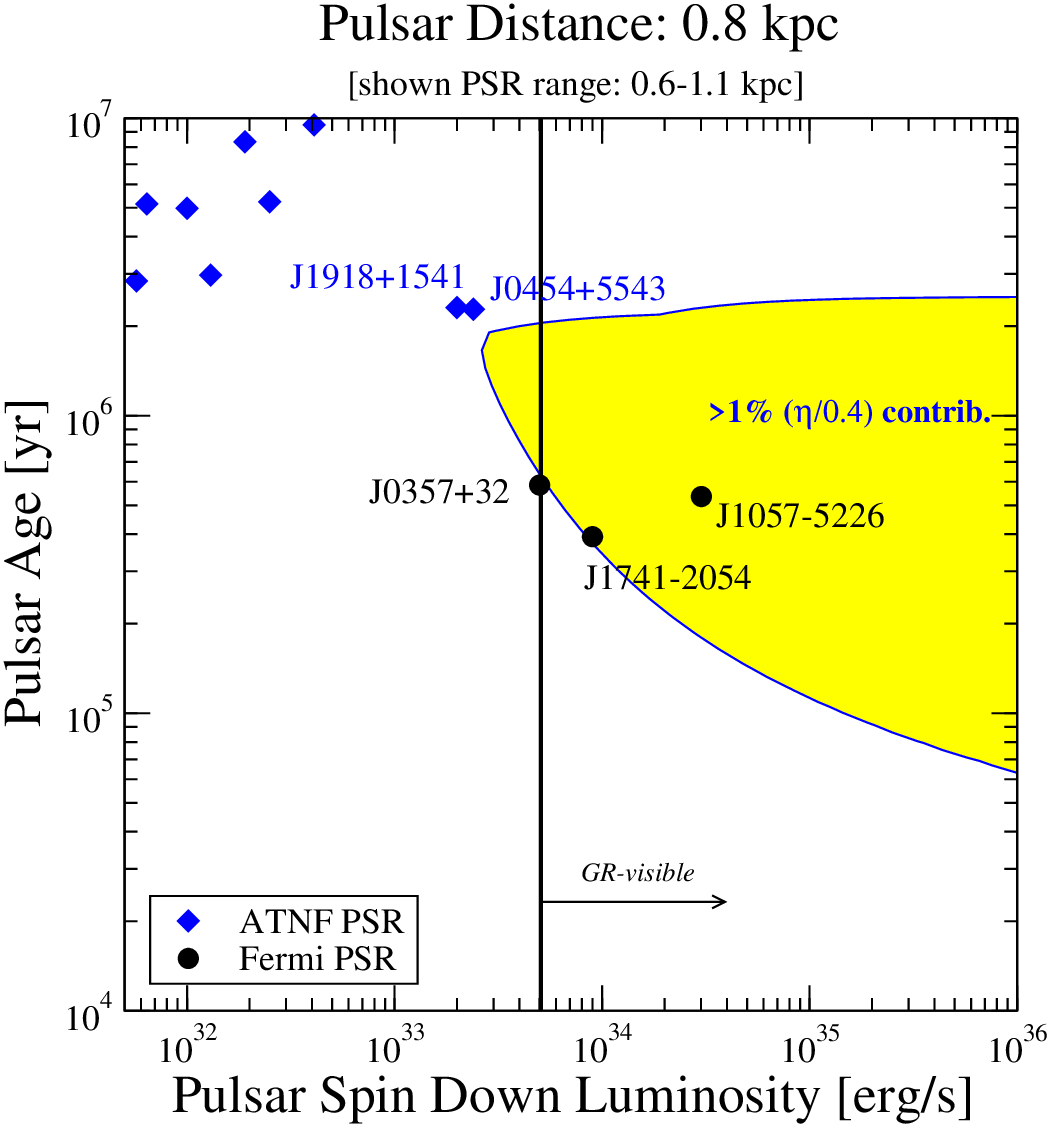}}\\
\caption{A comparison of the gamma-ray detectability of pulsars and of the predicted contribution to the local \epem flux, in the plane defined by the pulsars age versus spin down luminosity. In the left panel we focus on pulsars at a distance of 0.4 kpc, while on the right of 0.8 kpc. We also show, for reference, the location of actual pulsars, for distances between 0.2 and 0.6 kpc (left) and between 0.6 and 1.1 kpc (right). The black vertical lines correspond to the gamma-ray pulsars with the lowest spin-down luminosity. \label{fig:edotage}}
\end{center}
\end{figure}

In Fig.~\ref{fig:edotage} we consider the parameter space plane defined by the pulsar age versus the pulsar spin down luminosity, for pulsars with a distance around 0.4 kpc (left) and around 0.8 kpc (right). Pulsars within the yellow-shaded area are expected to contribute to the local \epem flux by more than 1\%, for an \epem efficiency $\eta=0.4$. We also include vertical black lines corresponding to the pulsars with the lowest spin-down luminosity (and hence lowest theoretical gamma-ray flux $\propto L_\gamma(\dot E)/D^2$) in that given distance range. This gives a benchmark of the Fermi-LAT gamma-ray sensitivity for pulsars in the given distance range (notice that this is only to guide the eye, since the gamma-ray flux depends on the pulsar-dependent gamma-ray efficiency and on the geometric flux correction factor).

We indicate, for purposes of illustration, pulsars from the ATNF catalogue (blue diamonds) and Fermi-LAT pulsars (black circles) with distances in the range between 0.2 and 0.6 kpc (left) and between 0.6 and 1.1 kpc (right). The corresponding yellow contours would of course be different for each individual pulsar at its own distance, but this illustrates, for instance, that the largest contribution to the local flux of \epem is expected to come from gamma-loud rather than from radio-loud pulsars (naturally, some gamma-ray pulsars expected to be significant contributors, such as Geminga, have also been detected at radio frequencies \cite{1997Natur.389..697M}). Also remarkably, regions of parameter space naively unaccessible to Fermi-LAT (i.e. to the left of the vertical lines) appear to be devoid of radio pulsars, implying that no significant \epem contributor is expected to be radio-loud and gamma-quiet.

The upper cutoff in age for the pulsars contributing to the local \epem flux simply stems from the maximal energy cutoff as a function of energy 
$$t\lesssim 2.3\times 10^6 \ {\rm yr}\left(\frac{1.4\times10^{-16}\ {\rm GeV}^{-1}{\rm s}^{-1}}{b_0}\right)\left(\frac{100\ {\rm GeV}}{E}\right).$$
The non-trivial dependence on pulsar age that cuts off the contribution from young pulsars depends, instead, on the fact that for young pulsars and for energies such that $E\ll E_{\rm max}=(b_0t)^{-1}$, the diffusion radius is
$$R_{\rm diff}\simeq0.5\ {\rm kpc}\left(\frac{D_0}{3.6\times 10^{28} {\rm cm}^2\ {\rm s}^{-1}}\frac{t}{10^5\ {\rm yr}}\right)^{1/2}\left(\frac{100\ {\rm GeV}}{E}\right)^{\delta/2},$$
so that the contribution from a young pulsar further away than 0.5 kpc is negligible at 100 GeV. In the picture where cosmic-ray \epem follow a random-walk path, this means that the diffusion radius of a young pulsar, at some given energy, is not large enough for the cosmic rays to diffuse to us. The shape of the parameter space shown in Fig.~\ref{fig:edotage} reproduces the analytic approximation outlined above.

\begin{figure}[t]
\begin{center}
\mbox{\includegraphics[width=8cm,clip]{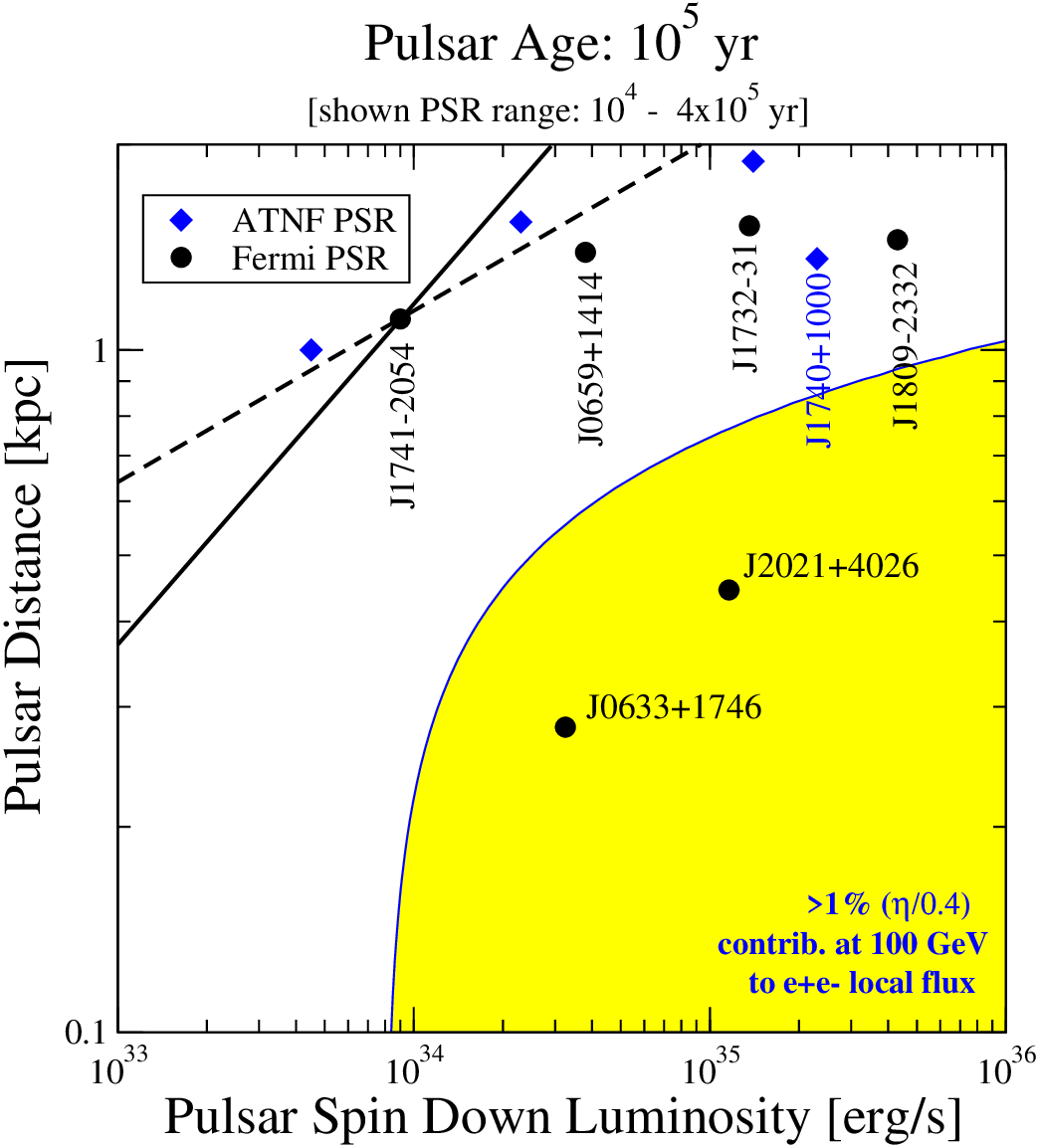}\qquad \includegraphics[width=8cm,clip]{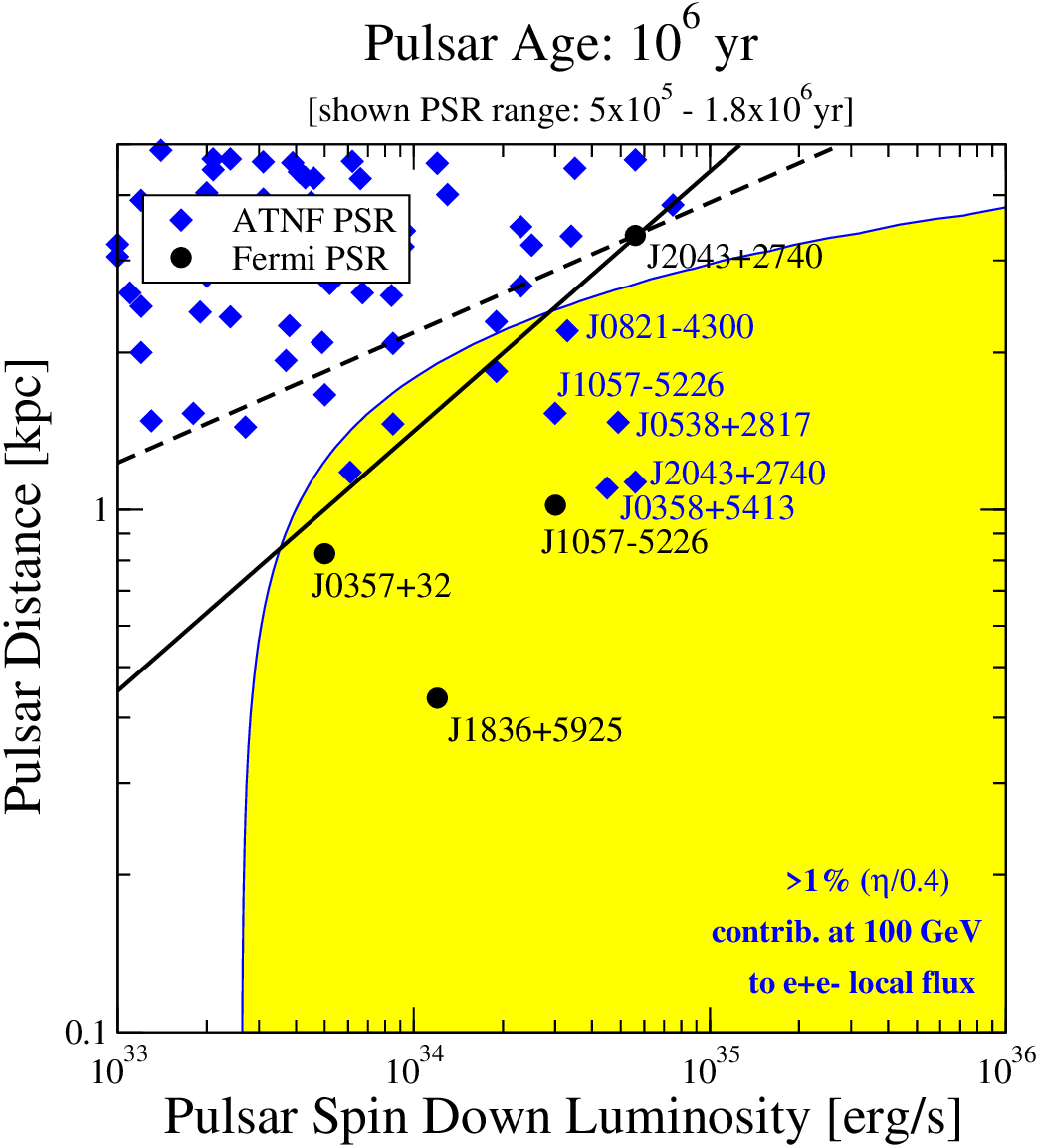}}\\
\caption{A comparison of the gamma-ray detectability of pulsars and of the predicted contribution to the local \epem flux, in the plane defined by the pulsar distance versus spin down luminosity. In the left panel we consider pulsars with an age of $10^5$ yr, while on the right of $10^6$ yr. We also show, for reference, the location of actual pulsars on the plane, for ages in the $0.1-4\times 10^5$ yr range (left) and in the $0.5-1.8\times 10^6$ yr range (right). To guide the eye, we draw lines corresponding to the gamma-ray pulsar with the lowest $L_\gamma/D^2$ in the range shown, for $L_\gamma=\dot E$ (solid) and $L_\gamma=\sqrt{10^{33}{\rm erg}\ {\rm s}^{-1}\times \dot E}$ (dashed). \label{fig:edotdist}}
\end{center}
\end{figure}

Fig.~\ref{fig:edotdist} shows the orthogonal parameter space direction of pulsar distance versus spin down luminosity, for pulsars with an age around $10^5$ yr (left) and $10^6$ yr (right). The black dots refer to  gamma-ray pulsars, while the blue diamonds to ATNF pulsars. For the former, we take the distance from the present analysis, while for the latter we use the distances quoted in the ATNF catalogue \cite{2005yCat.7245....0M}. Notice that we include all ATNF pulsars in the specified age ranges ($0.1-4\times 10^5$ yr in the left panel and  $0.5-1.8\times 10^6$ yr in the right panel), irrespective of whether or not they are theoretically expected to produce gamma rays. In the context of the outer gap model, for instance, the condition for a  rotation powered pulsar to produce gamma rays is that the ratio of the physical size of the outer gap over the ratio of the light cylinder be less than one, see e.g. Ref.~\cite{1994AIPC..304..116C} and \cite{2001A&A...368.1063Z}. This condition can be re-expressed in terms of the pulsar period $P$ in seconds and of the pulsar magnetic field at the neutron star surface $B_{12}$ in units of $10^{12}$ G as $$g=5.5\times P^{26/21}\ B_{12}^{-4/7}<1.$$ Not all the ATNF pulsars we show fulfill this criterion. As before, we add black lines to help guide the eye to the region of parameter space that should be detectable by Fermi. Here, we draw lines corresponding to the gamma-ray pulsar with the lowest $L_\gamma/D^2$ in the range shown, for $L_\gamma=\dot E$ (solid) and $L_\gamma=\sqrt{10^{33}{\rm erg}\ {\rm s}^{-1}\times \dot E}$ (dashed). Objects to the lower-right of the lines should give (for a gamma-ray efficiency and geometric flux correction factor comparable to the reference pulsar) a gamma-ray flux detectable by Fermi-LAT.

The shape of the parameter space where pulsars are expected to contribute to the local flux of \epem is here understood by considering the diffusion radius in the limit of mature enough pulsars, i.e. for energies such that $E/E_{\rm max}\simeq 1$. In this case, the diffusion radius is independent of the pulsar age, and it takes the value:
$$R_{\rm diff}\simeq2.7\ {\rm kpc}\left(\frac{D_0}{3.6\times 10^{28} {\rm cm}^2\ {\rm s}^{-1}}\frac{1.4\times 10^{-16}\ {\rm s}}{b_0}\right)^{1/2}\left(\frac{100\ {\rm GeV}}{E}\right)^{1-\delta}.$$
The distance above sets the quadratic-exponential decay scale for the local \epem flux with the pulsar distance $r$, $N_e\propto \exp[-(r/R_{\rm diff})^2]$ observed in the figure. The pulsar age influences the precise location of the contour in the plane via the effect on the \epem spectrum of the maximal energy exponential cutoff.

Fig.~\ref{fig:edotdist}, right, shows that pulsars which have not been detected in gamma rays yet (those corresponding to the blue diamonds) can in principle contribute significantly to the cosmic-ray flux as long as (i) they have a large enough spin-down luminosity and (ii) they are close enough (from the figure to the right we infer a distance cutoff around 2-3 kpc for ages on the order of a mega-year). We label some of the ATNF pulsars that follow in the portion of parameter space where pulsars are predicted to contribute to the percent level to the \epem flux, and that fulfill the above-mentioned gamma-ray pulsar outer-gap condition $g<1$. We find that several of these gamma-ray quiet ATNF pulsars (as of the first Fermi pulsar catalogues) can potentially give significant contributions to the \epem flux. Those same objects, however, are also expected to give a potentially detectable gamma-ray flux, and are thus ideal targets for LAT monitoring for a pulsed emission. Incidentally, we notice that the age of these pulsars, $t\sim10^6$ yr, indicates that their most significant contribution to the \epem flux would fall in the energy range around 100-200 GeV (see e.g. the inset in Fig.~\ref{fig:Rpeak}, right). 

Notice that contours of constant contribution from pulsars to the local \epem flux at 100 GeV would follow the shape of the yellow contours in Fig.~\ref{fig:edotdist}. This means that, for instance, the ATNF gamma-ray quiet pulsars close to the edge and above the solid black line in the right panel, while potentially beyond the Fermi gamma-ray sensitivity, are less likely to significantly contribute to the flux of \epem at Earth.

\section{Summary and Conclusions}\label{sec:concl}

In this paper we estimated the contribution to the local electron-positron flux from the gamma-ray pulsars in the first Fermi-LAT pulsar catalogue. We carried out estimates of the pulsar distances and their uncertainties from the Fermi gamma-ray flux and pulse shape. We calibrated our distance estimates with those pulsars with available alternate distance determinations. For a fixed choice of pulsar parameters relevant to the injection of electrons and positrons, we estimated the contribution of the set of known gamma-ray pulsars to the local high-energy cosmic-ray lepton flux. We found that ten Fermi-LAT pulsars can potentially contribute significantly to this flux. Of these ten potential contributors,  six are among newly discovered pulsars, found in the first Fermi blind-search campaign, and have very dim or no associated radio signal. We explored the ranges of pulsar ages and distances where we theoretically expect gamma-ray pulsars to contribute most significantly to the local cosmic-ray lepton flux. We found that regions of parameter space exist where gamma-ray pulsars would significantly contribute to the \epem flux without producing a gamma-ray flux detectable by Fermi-LAT. However, known radio pulsars fall in regions of parameter space where the \epem contribution is predicted to be typically much smaller than those regions where Fermi-LAT pulsars exist. We also found that radio pulsars with ages around one mega-year that have not yet been detected in gamma rays might also contribute significantly to the local \epem flux. In summary, we conclude that gamma-ray searches for mature and relatively nearby pulsars play a crucial role in understanding the origin of the local high-energy cosmic-ray electron and positron flux.

  
\section*{Acknowledgments}

SP is supported by an Outstanding Junior Investigator Award from the US Department of Energy (DoE), Office of Science, High Energy Physics, DoE Contract DEFG02-04ER41268. SP and LG are supported by NSF Grant PHY-0757911.

\section*{References}

\providecommand{\newblock}{}

\end{document}